\documentclass[11pt]{article}

\begin{document}
\title{On the Vacuum Energy in Expanding
Space-Times\thanks{Talk presented at the 18th IAP Colloquium in
Paris, July 1-5, 2002}}

\author{Achim Kempf\\
Department of Applied Mathematics\\ University of Waterloo,
Waterloo N2L 3G1, Canada}
\date{}

\maketitle

\begin{abstract}
If there is a shortest length in nature, for example at the Planck
scale of $10^{-35}m$, then the cosmic expansion should continually
create new comoving modes. A priori, each of the new modes comes
with its own vacuum energy, which could contribute to the
cosmological constant. I discuss possible mathematical models for
a shortest length and an explicit model for a corresponding
mode-generating mechanism.
\end{abstract}
\section{Introduction}
Theoretical evidence points towards the existence of a finite
shortest length in nature. In particular, general relativity and
quantum theory together indicate that the concept of distance
loses operational meaning at the Planck scale: In any
`microscope', test particles aimed to resolve a region of
Planckian size ($\approx 10^{-35}m$) would possess sufficient
momentum uncertainty to significantly randomly curve and thereby
disturb the region in space that they were meant to resolve.

The existence of a finite shortest length in nature could have
cosmological implications. This is because in an expanding
universe the independent modes are comoving modes, i.e. modes
whose wavelength expands with the universe. Therefore, if there is
a finite minimum length in nature, then new comoving modes of
initially Planckian wave length must be created continually. A
priori, each new mode comes with its own vacuum energy - which
could contribute to the cosmological `constant', and which could
also be related to the inflaton potential.

During inflation, in particular, the new modes expand only about
four or five orders of magnitude before their dynamics freezes
upon crossing the Hubble horizon \cite{infl}. After such a brief
expansion period the new modes could still possess properties that
can be traced back to Planck scale physics. Therefore, relics of
Planck scale physics which were frozen at horizon crossing could
later have contributed to the seeding of density perturbations and
may be observable in the CMB and in the subsequent structure
formation. A signature of Planck scale effects could exist, for
example, in contributions to the scalar/tensor ratio
\cite{kinney}. Crucial in this context is the curl component of
the polarization spectrum of the CMB, which is likely to be
measurable with MAP, Planck or a later satellite-based telescope.

\section{Models of space-time with an ultraviolet cutoff}

Any concrete candidate model of the short-distance structure of
space-time must explain how new comoving modes continually arise
and should therefore imply potentially measurable predictions for
the cosmological constant, the inflaton potential and the seeding
of structure formation.

A straightforward possibility is to model space-time as being
discrete. It is still unclear, however, how space-time's
continuity at large scales could emerge from a lattice
description, in particular, as space-time is expanding. It is
tempting, therefore, to speculate that a quantum gravity theory
such as M-theory, see \cite{pol}, or a foam theory, see
\cite{foam}, might eventually reveal the structure of space-time
as being in between discrete and continuous, perhaps such as to
combine the ultraviolet finiteness of lattices with continuous
symmetry properties of manifolds.

Concretely, let us begin by asking whether it is at all possible
that the points of space or of space-time form a set whose
cardinality is in between discrete (i.e. countable) and
continuous. The answer is a qualified no:

Cantor already conjectured (tragically) that there is no set whose
cardinality lies in between discrete and continuous. In the year
1900, this so-called continuum hypothesis (CH) was listed by
Hilbert as the first in his list of 23 problems for the 20th
century. By the middle of the 20th century the problem's subtle
solution was found by G{\"o}del and Cohen. They showed that,
within standard set theory, CH can neither be proved nor
disproved.

This result means that on the one hand we could enforce the
existence of sets of intermediate cardinality by simply claiming
their existence in a new axiom for set theory. On the other hand,
however, the answer of Cohen and G{\"o}del implies that it would
be impossible to explicitly construct any such set within standard
set theory, since otherwise CH could be disproved. This also means
that any \it explicit \rm set of space-time points can only be of
discrete or continuous cardinality (theoretically, sets of
space-time points could also be of higher than continuous
cardinality, but it appears safe to discard this possibility since
it would imply more rather than fewer ultraviolet divergencies).

We had asked whether there is a possibility by which nature might
combine the ultraviolet finiteness of lattices with the continuous
symmetry properties of manifolds. There still is a possibility.

\section{Finiteness of the density of degrees of freedom}

Let us recall that physical theories are formulated not directly
in terms of points in space or in space-time but rather in terms
of the functions over the set of points. This suggests a whole new
class of mathematical models for a finite minimum length, which
might beautifully combine the ultraviolet finiteness of lattices
with the continuous symmetry properties of manifolds:

As first proposed in \cite{ak-prl}, fields in space-time could be
functions over a continuous manifold as usual, while, crucially,
the class of fields that can occur is such that if a field is
sampled only at discrete points then its amplitudes can already be
reconstructed at \it all \rm points in the manifold - if the
sampling points are spaced densely enough. The maximum average
sample spacing which allows one to reconstruct the continuous
field from discrete samples could be on the order of the Planck
scale.

Since any one of all sufficiently tightly spaced lattices would
allow reconstruction, no particular lattice would be preferred.
Thus, all the continuity and symmetry properties of the manifold
would be preserved. The physical theory could be written,
equivalently, either as living on a continuous manifold, thereby
displaying all its external symmetries, or as living on any one of
the sampling lattices of sufficiently small average spacing,
thereby displaying its ultraviolet finiteness. Physical fields,
while being continuous or even differentiable, would possess only
a finite density of degrees of freedom.

The mathematics of classes of functions which can be reconstructed
from discrete samples is well-known, namely as \it sampling
theory, \rm in the information theory community, where it plays a
central role in the theory of sources and channels of continuous
information as developed by Shannon, see \cite{shannon}.
\section{Sampling theory}
The simplest example in sampling theory is the Shannon sampling
theorem: Choose a frequency $\omega_{max}$. Consider the class
$B_{\omega_{max}}$ of continuous functions $f$ whose frequency
content is limited to the interval $(-\omega_{max},\omega_{max})$,
i.e. for which:
%\begin{equation}
$ \tilde{f}(\omega)~=~\int_{-\infty}^{\infty} f(x) e^{-i\omega x}
= 0 \mbox{~~~whenever ~~} \vert\omega\vert\ge\omega_{max}.$
%\end{equation}
If the amplitudes $f(x_n)$ of such a function are known at
equidistantly spaced discrete values $\{x_n\}$ whose spacing is
$\pi/\omega_{max}$ or smaller, then the function's  amplitudes
$f(x)$ can be reconstructed for all $x$. The reconstruction
formula is:
\begin{equation}
f(x)~ =~ \sum_{n=-\infty}^\infty ~f(x_n) ~
\frac{\sin[(x-x_n)\omega_{max}]}{(x-x_n)\omega_{max}}
\end{equation}
For the proof, note that $\tilde{f}$ has compact support and can
therefore be represented both as a Fourier series of the $f(x_n)$
and as a Fourier transform of $f(x)$. The theorem is in ubiquitous
use for example in digital audio and video. Sampling theory, see
\cite{ferreira}, studies generalizations of the theorem for
various different classes of functions, for non-equidistant
sampling, for multi-variable functions and it investigates the
effect of noise, which could be quantum fluctuations in our case.
Due to its engineering origin, sampling theory for generic pseudo-
Riemannian manifolds is still virtually undeveloped, but should be
of great interest. As was shown in \cite{ak-prl}, generalized
sampling theorems automatically arise from stringy uncertainty
relations, namely whenever there is a finite minimum position
uncertainty $\Delta x_{min}$, as e.g. in uncertainty relations of
the type: $\Delta x \Delta p \ge \frac{\hbar}{2}(1 + \beta (\Delta
p)^2 +...)$, see \cite{ucr}. A few technical remarks: the
underlying mathematics is that of symmetric non self-adjoint
operators. Through a theorem of Naimark, unsharp variables of POVM
type arise as special cases.
\section{Application in Cosmology}
It is possible to obtain explicit candidate models of quantum
field theory close to the Planck scale by implementing a minimum
length in nature as a sampling theoretical maximum information
density. To this end, in \cite{cosm}, the above mentioned stringy
uncertainty relations were implemented through their corresponding
commutation relations. Physical fields then automatically become
reconstructible everywhere from their samples on any lattice of
spacing smaller than the Planck length.

The calculations were done for flat expanding background
space-times, such as de Sitter space. As expected, comoving modes
are continually being created. The modes' wave equations are still
of the form $\Phi^{\prime\prime} + m(\eta) \Phi^{\prime} + n(\eta)
\phi=0$ but each mode now possesses of course a starting time
$\eta_0$. At a given mode's starting time $\eta_0$ the coefficient
functions $m(\eta)$ and $n(\eta)$ are singular. This singularity
has not yet been fully understood. However, since it determines
the initial conditions and the vacuum its understanding will be
crucial for predicting the size and signature of potentially
observable effects. For the Hamiltonian, mode generation happens
by continually adding the new modes' $a_{\vec{k}}$ and
$a_{\vec{k}}^\dagger$ operators:
$$
H = \int_{\vec{k}^2 <
\frac{a(\eta)^2}{l^2_{\tiny{Pl}}}}~~d^3k~~\left(
\Phi^\prime_{\vec{k}}\Pi_{\vec{k}} - {\cal{L}}_{\vec{k}}\right)
$$
Here, $a(\eta)$ is the scale factor. Clearly, there is a continual
production of vacuum energy if it is not artificially normal
ordered away. Its size and properties are currently being
investigated. It should be very interesting to apply sampling
theory also to holographic bounds on the area density of
information.

\end{document}